Hellenic Complex Systems Laboratory

# Calculation of the confidence bounds for the fraction nonconforming of normal populations of measurements in clinical laboratory medicine

Technical Report IV

Aristides T. Hatjimihail
2019

# 1. Abstract

The fraction nonconforming is a key quality measure used in statistical quality control design in clinical laboratory medicine. The confidence bounds of normal populations of measurements for the fraction nonconforming each of the lower and upper quality specification limits when both the random and the systematic error are unknown can be calculated using the noncentral t-distribution, as it is described in detail and illustrated with examples.

# 2. Introduction

The quality control (QC) in clinical laboratory medicine is based mainly on the control charts for variables. Although the fraction nonconforming is used in the control charts for attributes, it is also a useful quality measure related to various process capability indices (Pearn & Kotz 2006). It is defined as "the ratio of the items nonconforming to the total number of items in a subgroup" (Juran et al. 1999).

The quality control design process (Hatjimihail 1992) is based on the quality specifications of the maximum total allowable analytical error, the upper acceptable bound for the fraction nonconforming, the lower acceptable bounds for the probabilities for the critical random and systematic error detection, and the upper acceptable bound for the false rejection (CLSI 2012; Linnet 1989).

It includes the following steps:

1. Estimation of the error of the measurement procedure based on a sample of $n$ control measurements.

2. Calculation of the critical errors of the measurement procedure based on an upper acceptable bound for the expected fraction nonconforming the maximum total allowable analytical error quality specification. This bound is usually set equal to 0.10.
3. Definition of the decision bounds for the quality control rules based on the lower acceptable bounds for the probabilities for the critical errors detection and the upper acceptable bound for false rejection.

These steps are performed at typically one to three levels of the measurand.

Therefore, the expected fraction nonconforming is a key measure of the quality control design process. While this measure is routinely calculated in the clinical laboratories, very little attention has been given to its confidence bounds.

In general, for the quality control estimations in laboratory medicine it is assumed that the error of a measurement procedure is normally distributed with known mean and standard deviation. Usually in practice we only know the mean $\overline{X}$ and the standard deviation $s$ of a sample of $n$ control measurements at each of various levels of the measurand.

As both the systematic and the random error of the measurement procedure are unknown lower and upper bounds for the tail probabilities of the lower ( $x_{LSL}$ ), and the upper ( $x_{USL}$ ) quality specifications of the measurement $X$ can be calculated with confidence level $\gamma$ using the noncentral $t$-distribution (Anonymous, 1987) , as described in detail in the Materials and Methods section.

Therefore, the purpose of this technical report is to introduce the noncentral t-distribution into clinical laboratory medicine and to describe its application for the calculation of the confidence bounds of the fraction nonconforming each of the lower and upper quality specification limits of populations of normal measurements in a clinical laboratory setting..

## 3. Materials and Methods

If *Z* a standard normal *N*(0, 1) deviate and *V* a chi-square deviate with *v* degrees of freedom, the distribution of the ratio is the noncentral *t*-distribution with *v* degrees of freedom and noncentrality parameter *δ*.

$$T_{\nu,\delta} = \frac{Z+\delta}{\left(\frac{V_\nu}{\nu}\right)^{-1/2}} \quad (0.1)$$

The cumulative density function of a noncentral *t*-distribution with *v* degrees of freedom and noncentrality parameter *δ* is defined as (Owen 1965):

$$F_{\nu,\delta}(t) = \frac{\sqrt{2\pi}}{\Gamma(\nu/2)2^{(\nu-2)/2}\sqrt{\pi}} \int_0^\infty \Phi\left(\frac{tx}{\sqrt{\nu}} - \delta\right) x^{\nu-1}\phi(x)dx \quad (0.2)$$

where $\Gamma(x)$ the gamma function and:

$$\varphi(x) = \frac{1}{\sqrt{2\pi}} e^{\frac{x^2}{2}} \quad (0.3)$$

$$\Phi(x) = \int_{-\infty}^{x} \varphi(t)dt \quad (0.4)$$

Scholz (Scholz 1994) presents an elegant derivation of the lower and upper bounds for the lower tail probability $p(x_{LSL})$ of a normal population with unknown mean *μ* and unknown standard deviation *σ*.

Briefly,

$$p(x_{LSL}) = p(x_{LSL}, \mu, \sigma) = P_{\mu,\sigma}(X \le x_{LSL}) = \Phi\left(\frac{x_{LSL}-\mu}{\sigma}\right) \quad (0.5)$$

If we denote with $\hat{p}$ an upper bound for $p(x_{LSL})$ with confidence level *γ* then

$$P_{\mu,\sigma}\left(\hat{p}_{UL}(\gamma) \ge p(x_{LSL})\right) = \gamma \quad (0.6)$$

Let

$$Z = \sqrt{n}\left(\mu - \bar{X}\right)/\sigma \quad (0.7)$$

a standard normal deviate, and

$$V = (n-1)s^2/\sigma^2 \quad (0.8)$$

a chi-square deviate with *n*-1 degrees of freedom.

If

$$\delta = \frac{\sqrt{n}\left(x_{LSL}-\mu\right)}{\sigma} = \sqrt{n}\ \Phi^{-1}\left(p\left(x_{LSL}\right)\right) \quad (0.9)$$

then,

$$\frac{\sqrt{n}\left(x_{LSL}-\bar{x}\right)}{s}=\frac{\frac{\sqrt{n}\left(x_{LSL}-\mu\right)}{\sigma}+\frac{\sqrt{n}\left(\mu-\bar{X}\right)}{\sigma}}{\frac{s}{\sigma}}=$$

$$\frac{\frac{\sqrt{n}\left(x_{LSL}-\mu\right)}{\sigma}+\frac{\sqrt{n}\left(\mu-\bar{X}\right)}{\sigma}}{\sqrt{\frac{\left(\frac{(n-1)s^2}{\sigma^2}\right)}{n-1}}}=\frac{Z+\delta}{\sqrt{\frac{V}{n-1}}}=T_{n-1,\delta} \tag{0.10}$$

where the $T_{n-1,\delta}$ has by definition a noncentral *t*-distribution with *n*-1 degrees of freedom noncentrality parameter $\delta$. According to the probability integral transformation (Queesenberry 2006) the random variable $U$ is uniformly distributed over (0,1). Therefore,

$$U=F_{n-1,\delta}\left(\frac{\sqrt{n}\left(x_{LSL}-\bar{X}\right)}{S}\right)=F_{n-1,\delta}(T_{n-1,\delta}) \tag{0.11}$$

$$\gamma=P(U\geq 1-\gamma) \tag{0.12}$$

As $F_{n-1,\delta}(x)$ is a decreasing function of $\delta$ we have

$$F_{n-1,\delta}\left(\frac{\sqrt{n}\left(x_{LSL}-\bar{X}\right)}{s}\right)\geq 1-\gamma \Leftrightarrow \delta\leq\hat{\delta} \tag{0.13}$$

where $\hat{\delta}$ is estimated by the following equation:

$$F_{n-1,\hat{\delta}}\left(\frac{\sqrt{n}\left(x_{LSL}-\bar{X}\right)}{s}\right)=1-\gamma \tag{0.14}$$

Consequently, the upper bound with confidence level $\gamma$ for the fraction nonconforming the lower quality specification limit is calculated as

$$\hat{p}_{UL}\left(\gamma\right)=\Phi\left(\frac{\hat{\delta}}{\sqrt{n}}\right) \tag{0.15}$$

As

$$P_{\mu,\sigma}\left(p_{UL}\left(\gamma\right)\geq p\left(x_{LSL}\right)\right)=\gamma \Leftrightarrow P_{\mu,\sigma}\left(\hat{p}_{UL}\left(\gamma\right)\leq p\left(x_{LSL}\right)\right)=1-\gamma \tag{0.16}$$

the lower bound with confidence level $\gamma$ for the fraction nonconforming the lower quality specification limit is calculated as

$$\hat{p}_{LL}\left(\gamma\right)=\hat{p}_{UL}\left(1-\gamma\right) \tag{0.17}$$

In addition, if we denote with $\hat{p}_{UU}\left(\gamma\right)$ the upper bound with confidence level $\gamma$ for the fraction nonconforming the upper quality specification limit then

$$\begin{gathered}P_{\mu,\sigma}\left(1-\hat{p}_{UU}\left(\gamma\right)\leq 1-p\left(x_{USL}\right)\right)=\gamma \Leftrightarrow \\ P_{\mu,\sigma}\left(1-\hat{p}_{UU}\left(\gamma\right)\geq 1-p\left(x_{USL}\right)\right)=1-\gamma\end{gathered} \tag{0.18}$$

Therefore, the lower and upper bounds with confidence level $\gamma$ for the fraction nonconforming the upper quality specification limit are calculated respectively as

$$\hat{p}'_{LU}(\gamma) = 1 - \hat{p}_{UU}(\gamma) \tag{0.19}$$

and

$$\hat{p}'_{UU}(\gamma) = 1 - \hat{p}_{UU}(1-\gamma) \tag{0.20}$$

Consequently, the lower and the upper bounds for the fraction nonconforming with confidence level $\gamma$ are estimated respectively as

$$\hat{fnc}_L(\gamma) = \hat{p}_{LL}\left(\frac{1+\gamma}{2}\right) + \hat{p}'_{LU}\left(\frac{1+\gamma}{2}\right) = \hat{p}_{UL}\left(\frac{1-\gamma}{2}\right) + 1 - \hat{p}_{UU}\left(\frac{1+\gamma}{2}\right) \tag{0.21}$$

and

$$\hat{fnc}_U(\gamma) = \hat{p}_{UL}\left(\frac{1+\gamma}{2}\right) + \hat{p}'_{UU}\left(\frac{1+\gamma}{2}\right) = \hat{p}_{UL}\left(\frac{1+\gamma}{2}\right) + 1 - \hat{p}_{UU}\left(\frac{1-\gamma}{2}\right) \tag{0.22}$$

If we denote with $c$ the true value of the measurand of the control material, and $e_{\max}$ the maximum total clinically allowable analytical error (Linnet 1989), expressed as a ratio to the true value, then

$$x_{LSL} = c(1-e_{\max}) \tag{0.23}$$

and

$$x_{USL} = c(1+e_{\max}) \tag{0.24}$$

To illustrate the application of the noncentral t-distribution for the calculation of the confidence bounds of the fraction nonconforming of normal populations of measurements in a clinical laboratory setting I calculated upper and lower confidence bounds for the fraction nonconforming the upper and lower quality specification limits of populations of normal measurements, assuming unknown systematic and random errors of the measurement procedure. The calculations were performed by MATLAB©, ver. 2018b, using the method described by Owen (Owen 1965).

It was assumed that a reference control material with a measurand with an assigned value of 100.00 units has been measured repeatedly by the measuring procedure and that the sample mean and standard deviation of the measurand were estimated from samples of 20 and 40 control measurements. It was also assumed that both the systematic and the random error of the measurement procedure were unknown and that the uncertainty of the assigned value of the measurand of the control material was negligible. The lower and upper confidence bounds for the fraction nonconforming each of the lower and upper quality specifications of the population of the measurements were calculated by the noncentral *t*-distribution as described previously.

The parameters of these calculations are presented in the Table 1.

| **figures** | | 1-2 | 3-4 | 5-6 | 7-8 |
|---|---|---|---|---|---|
| **tables** | | 2 | 3 | 4 | 5 |
| **assigned value of the measurand of the control material** | $\mu$ | 100.00 | 100.00 | 100.00 | 100.00 |
| **sample mean** | $\bar{X}$ | 100.00-106.00 | 100.00-106.00 | 100.00-106.00 | 100.00-106.00 |
| **increments of the sample mean** | | 0.10 | 0.10 | 0.10 | 0.10 |
| **sample s.d.** | $s$ | 0.10-6.00 | 0.10-6.00 | 0.10-6.00 | 0.10-6.00 |
| **increments of the sample s.d.** | | 0.10 | 0.10 | 0.10 | 0.10 |
| **total allowable analytical error** | $e_{max}$ | 0.10 | 0.10 | 0.20 | 0.20 |
| **confidence level** | $\gamma$ | 0.95 | 0.95 | 0.95 | 0.95 |
| **sample size** | $n$ | 20 | 40 | 20 | 40 |

**Table 1:** *The parameters of the calculations of the confidence bounds for the fraction nonconforming of the populations of the normal control measurements.*

# 4. Results

The results of the study are described in the following subsections 4.1 and 4.2. They are presented in the tables 2-5 and in the figures 1- 8. They are also presented in the respective MATLAB© figures included in the supporting information file MATLAB_Figures.zip so that the data can be explored interactively in the MATLAB© environment.

## 4.1. Fraction nonconforming the lower quality specification limit

Figures 1, 3, 5, and 7 present the expected fraction nonconforming the lower quality specification limit (surface II) and the lower (surface III) and upper (surface I) confidence bounds for the fraction nonconforming at a 95% confidence level versus the sample mean and the sample standard deviation. The confidence bounds were calculated using the noncentral *t*-distribution. It was assumed a known true value of the measurand of the control material equal to 100.00 measurement units. For the figures 1 and 3 it was assumed a total allowable analytical error equal to 0.10 while for the figures 5 and 7 it was assumed a total allowable analytical error equal to 0.20. The respective samples were samples of 20 control measurements for the figures 1 and 5, and samples of 40 control measurements for the figures 3 and 7.

## 4.2. Fraction nonconforming the upper quality specification limit

Figures 2, 4, 6, and 8 present the expected fraction nonconforming the upper quality specification limit (surface II) and the lower (surface III) and upper (surface I) confidence bounds for the fraction nonconforming at a 95% confidence level versus the sample mean and the sample standard deviation. The confidence bounds were calculated using the noncentral *t*-distribution. It was assumed a known true value of the measurand of the control material equal to 100.00 measurement units. For the figures 2 and 4 it was assumed a total allowable analytical error equal to 0.10 while for the figures 6 and 8 it was assumed a total allowable analytical error equal to 0.20. The respective samples were samples of 20 control measurements for the figures 2 and 6, and samples of 40 control measurements for the figures 4 and 8.

The results are summarized in the tables 2 and 3, assuming a total allowable analytical error equal to 0.10, and in the tables 4 and 5, assuming a total allowable analytical error equal to 0.20. The respective samples were samples of 20 control measurements for the tables 2 and 4, and samples of 40 control measurements for the tables 3 and 5.

The confidence bounds increase with the estimated systematic and random error of the measurements and decrease with the sample size and the total allowable analytical error.

Although there have been assumed samples with a positive systematic error, it can be easily shown that the bounds for the fraction nonconforming the upper quality specification limit for a systematic error *se* are equal to the respective bounds for the fraction nonconforming the lower quality specification limit for a systematic error –*se*.

| confidence bounds for the fraction nonconforming at a 95% confidence level | | | | | | | |
|---|---|---|---|---|---|---|---|
| size of the control measurements sample $n$ = 20 | | | | | | | |
| total allowable analytical error *taae* = 0.10 | | | | | | | |
| **parameters of the sample** | | **fraction nonconforming the lower quality specification limit** | | | **fraction nonconforming the upper quality specification limit** | | |
| $\bar{X}$ | $s$ | **expected** | **lower bound** | **upper bound** | **expected** | **lower bound** | **upper bound** |
| **100.00** | **1.00** | 7,62E-24 | 8,23E-37 | 1,79E-13 | 7,62E-24 | 8,23E-37 | 1,79E-13 |
| | **2.00** | 2,87E-07 | 1,11E-10 | 1,62E-04 | 2,87E-07 | 1,11E-10 | 1,62E-04 |
| | **3.00** | 4,29E-04 | 9,84E-06 | 9,26E-03 | 4,29E-04 | 9,84E-06 | 9,26E-03 |
| | **4.00** | 6,21E-03 | 5,98E-04 | 4,23E-02 | 6,21E-03 | 5,98E-04 | 4,23E-02 |
| | **5.00** | 2,28E-02 | 4,28E-03 | 9,02E-02 | 2,28E-02 | 4,28E-03 | 9,02E-02 |
| **101.00** | **1.00** | 1,91E-28 | 4,36E-44 | 6,10E-16 | 1,13E-19 | 3,21E-30 | 3,13E-11 |
| | **2.00** | 1,90E-08 | 1,58E-12 | 3,67E-05 | 3,40E-06 | 5,31E-09 | 6,29E-04 |
| | **3.00** | 1,23E-04 | 1,42E-06 | 4,60E-03 | 1,35E-03 | 5,77E-05 | 1,77E-02 |
| | **4.00** | 2,98E-03 | 1,95E-04 | 2,78E-02 | 1,22E-02 | 1,67E-03 | 6,27E-02 |
| | **5.00** | 1,39E-02 | 2,04E-03 | 6,76E-02 | 3,59E-02 | 8,48E-03 | 1,18E-01 |
| **102.00** | **1.00** | 1,78E-33 | 4,77E-52 | 1,23E-18 | 6,22E-16 | 2,59E-24 | 3,23E-09 |
| | **2.00** | 9,87E-10 | 1,51E-14 | 7,32E-06 | 3,17E-05 | 1,72E-07 | 2,16E-03 |
| | **3.00** | 3,17E-05 | 1,72E-07 | 2,16E-03 | 3,83E-03 | 2,86E-04 | 3,21E-02 |
| | **4.00** | 1,35E-03 | 5,77E-05 | 1,77E-02 | 2,28E-02 | 4,28E-03 | 9,02E-02 |
| | **5.00** | 8,20E-03 | 9,13E-04 | 4,97E-02 | 5,48E-02 | 1,59E-02 | 1,53E-01 |
| **103.00** | **1.00** | 6,12E-39 | 1,08E-60 | 1,46E-21 | 1,28E-12 | 4,33E-19 | 1,99E-07 |
| | **2.00** | 4,02E-11 | 9,85E-17 | 1,29E-06 | 2,33E-04 | 3,82E-06 | 6,57E-03 |
| | **3.00** | 7,34E-06 | 1,77E-08 | 9,63E-04 | 9,82E-03 | 1,20E-03 | 5,52E-02 |
| | **4.00** | 5,77E-04 | 1,56E-05 | 1,09E-02 | 4,01E-02 | 9,97E-03 | 1,26E-01 |
| | **5.00** | 4,66E-03 | 3,86E-04 | 3,59E-02 | 8,08E-02 | 2,81E-02 | 1,94E-01 |
| **104.00** | **1.00** | 7,79E-45 | 4,99E-70 | 1,03E-24 | 9,87E-10 | 1,51E-14 | 7,32E-06 |
| | **2.00** | 1,28E-12 | 4,33E-19 | 1,99E-07 | 1,35E-03 | 5,77E-05 | 1,77E-02 |
| | **3.00** | 1,53E-06 | 1,53E-09 | 4,06E-04 | 2,28E-02 | 4,28E-03 | 9,02E-02 |
| | **4.00** | 2,33E-04 | 3,82E-06 | 6,57E-03 | 6,68E-02 | 2,13E-02 | 1,72E-01 |
| | **5.00** | 2,56E-03 | 1,54E-04 | 2,54E-02 | 1,15E-01 | 4,70E-02 | 2,41E-01 |
| **105.00** | **1.00** | 3,67E-51 | 4,76E-80 | 4,25E-28 | 2,87E-07 | 1,11E-10 | 1,62E-04 |
| | **2.00** | 3,19E-14 | 1,29E-21 | 2,71E-08 | 6,21E-03 | 5,98E-04 | 4,23E-02 |
| | **3.00** | 2,87E-07 | 1,11E-10 | 1,62E-04 | 4,78E-02 | 1,30E-02 | 1,41E-01 |
| | **4.00** | 8,84E-05 | 8,51E-07 | 3,83E-03 | 1,06E-01 | 4,16E-02 | 2,29E-01 |
| | **5.00** | 1,35E-03 | 5,77E-05 | 1,77E-02 | 1,59E-01 | 7,45E-02 | 2,96E-01 |

**Table 2:** *Confidence bounds for the fraction nonconforming the lower and upper quality specification limits assuming a true value of the measurand of the control material equal to 100.00 measurement units, a sample of 20 control measurements, and a total allowable analytical error equal to 0.10.*

| confidence bounds for the fraction nonconforming at a 95% confidence level | | | | | | | |
|---|---|---|---|---|---|---|---|
| size of the control measurements sample $n$ = 40 | | | | | | | |
| total allowable analytical error *taae* = 0.10 | | | | | | | |
| **parameters of the sample** | | **fraction nonconforming the lower quality specification limit** | | | **fraction nonconforming the upper quality specification limit** | | |
| $\overline{X}$ | $s$ | **expected** | **lower bound** | **upper bound** | **expected** | **lower bound** | **upper bound** |
| **100.00** | **1.00** | 7,62E-24 | 1,12E-32 | 2,79E-16 | 7,62E-24 | 1,12E-32 | 2,79E-16 |
| | **2.00** | 2,87E-07 | 1,35E-09 | 2,89E-05 | 2,87E-07 | 1,35E-09 | 2,89E-05 |
| | **3.00** | 4,29E-04 | 3,25E-05 | 4,01E-03 | 4,29E-04 | 3,25E-05 | 4,01E-03 |
| | **4.00** | 6,21E-03 | 1,25E-03 | 2,50E-02 | 6,21E-03 | 1,25E-03 | 2,50E-02 |
| | **5.00** | 2,28E-02 | 7,23E-03 | 6,19E-02 | 2,28E-02 | 7,23E-03 | 6,19E-02 |
| **101.00** | **1.00** | 1,91E-28 | 4,24E-39 | 2,52E-19 | 1,13E-19 | 7,37E-27 | 1,62E-13 |
| | **2.00** | 1,90E-08 | 3,13E-11 | 4,70E-06 | 3,40E-06 | 4,13E-08 | 1,52E-04 |
| | **3.00** | 1,23E-04 | 5,84E-06 | 1,71E-03 | 1,35E-03 | 1,56E-04 | 8,77E-03 |
| | **4.00** | 2,98E-03 | 4,61E-04 | 1,51E-02 | 1,22E-02 | 3,13E-03 | 4,01E-02 |
| | **5.00** | 1,39E-02 | 3,73E-03 | 4,39E-02 | 3,59E-02 | 1,33E-02 | 8,55E-02 |
| **102.00** | **1.00** | 1,78E-33 | 4,00E-46 | 1,18E-22 | 6,22E-16 | 1,21E-21 | 4,90E-11 |
| | **2.00** | 9,87E-10 | 5,15E-13 | 6,52E-07 | 3,17E-05 | 9,02E-07 | 6,85E-04 |
| | **3.00** | 3,17E-05 | 9,02E-07 | 6,85E-04 | 3,83E-03 | 6,49E-04 | 1,79E-02 |
| | **4.00** | 1,35E-03 | 1,56E-04 | 8,77E-03 | 2,28E-02 | 7,23E-03 | 6,19E-02 |
| | **5.00** | 8,20E-03 | 1,82E-03 | 3,04E-02 | 5,48E-02 | 2,34E-02 | 1,15E-01 |
| **103.00** | **1.00** | 6,12E-39 | 9,36E-54 | 2,90E-26 | 1,28E-12 | 4,98E-17 | 7,79E-09 |
| | **2.00** | 4,02E-11 | 6,02E-15 | 7,72E-08 | 2,33E-04 | 1,40E-05 | 2,64E-03 |
| | **3.00** | 7,34E-06 | 1,20E-07 | 2,56E-04 | 9,82E-03 | 2,33E-03 | 3,44E-02 |
| | **4.00** | 5,77E-04 | 4,88E-05 | 4,91E-03 | 4,01E-02 | 1,54E-02 | 9,23E-02 |
| | **5.00** | 4,66E-03 | 8,47E-04 | 2,05E-02 | 8,08E-02 | 3,91E-02 | 1,52E-01 |
| **104.00** | **1.00** | 7,79E-45 | 5,44E-62 | 3,70E-30 | 9,87E-10 | 5,15E-13 | 6,52E-07 |
| | **2.00** | 1,28E-12 | 4,98E-17 | 7,79E-09 | 1,35E-03 | 1,56E-04 | 8,77E-03 |
| | **3.00** | 1,53E-06 | 1,37E-08 | 8,90E-05 | 2,28E-02 | 7,23E-03 | 6,19E-02 |
| | **4.00** | 2,33E-04 | 1,40E-05 | 2,64E-03 | 6,68E-02 | 3,04E-02 | 1,33E-01 |
| | **5.00** | 2,56E-03 | 3,74E-04 | 1,36E-02 | 1,15E-01 | 6,22E-02 | 1,97E-01 |
| **105.00** | **1.00** | 3,67E-51 | 7,83E-71 | 2,45E-34 | 2,87E-07 | 1,35E-09 | 2,89E-05 |
| | **2.00** | 3,19E-14 | 2,92E-19 | 6,69E-10 | 6,21E-03 | 1,25E-03 | 2,50E-02 |
| | **3.00** | 2,87E-07 | 1,35E-09 | 2,89E-05 | 4,78E-02 | 1,95E-02 | 1,05E-01 |
| | **4.00** | 8,84E-05 | 3,71E-06 | 1,37E-03 | 1,06E-01 | 5,56E-02 | 1,85E-01 |
| | **5.00** | 1,35E-03 | 1,56E-04 | 8,77E-03 | 1,59E-01 | 9,44E-02 | 2,50E-01 |

**Table 3:** *Confidence bounds for the fraction nonconforming the lower and upper quality specification limits assuming a true value of the measurand of the control material equal to 100.00 measurement units, a sample of 40 control measurements, and a total allowable analytical error equal to 0.10.*

| confidence bounds for the fraction nonconforming at a 95% confidence level | | | | | | | |
|---|---|---|---|---|---|---|---|
| size of the control measurements sample *n* = 20 | | | | | | | |
| total allowable analytical error *taae* = 0.20 | | | | | | | |
| **parameters of the sample** | | **fraction nonconforming the lower quality specification limit** | | | **fraction nonconforming the upper quality specification limit** | | |
| $\overline{X}$ | $s$ | **expected** | **lower bound** | **upper bound** | **expected** | **lower bound** | **upper bound** |
| **100.00** | **1.00** | 2,75E-89 | 1,84E-140 | 1,86E-48 | 2,75E-89 | 1,84E-140 | 1,86E-48 |
| | **2.00** | 7,62E-24 | 8,23E-37 | 1,79E-13 | 7,62E-24 | 8,23E-37 | 1,79E-13 |
| | **3.00** | 1,31E-11 | 1,69E-17 | 7,01E-07 | 1,31E-11 | 1,69E-17 | 7,01E-07 |
| | **4.00** | 2,87E-07 | 1,11E-10 | 1,62E-04 | 2,87E-07 | 1,11E-10 | 1,62E-04 |
| | **5.00** | 3,17E-05 | 1,72E-07 | 2,16E-03 | 3,17E-05 | 1,72E-07 | 2,16E-03 |
| **101.00** | **1.00** | 3,28E-98 | 1,32E-154 | 3,23E-53 | 8,53E-81 | 5,29E-127 | 6,34E-44 |
| | **2.00** | 4,32E-26 | 2,31E-40 | 1,12E-14 | 1,05E-21 | 1,98E-33 | 2,53E-12 |
| | **3.00** | 1,28E-12 | 4,33E-19 | 1,99E-07 | 1,20E-10 | 5,51E-16 | 2,33E-06 |
| | **4.00** | 7,60E-08 | 1,39E-11 | 7,83E-05 | 1,02E-06 | 8,06E-10 | 3,24E-04 |
| | **5.00** | 1,33E-05 | 4,49E-08 | 1,34E-03 | 7,23E-05 | 6,23E-07 | 3,42E-03 |
| **102.00** | **1.00** | 1,44E-107 | 1,93E-169 | 3,29E-58 | 9,74E-73 | 3,11E-114 | 1,27E-39 |
| | **2.00** | 1,91E-28 | 4,36E-44 | 6,10E-16 | 1,13E-19 | 3,21E-30 | 3,13E-11 |
| | **3.00** | 1,12E-13 | 9,36E-21 | 5,34E-08 | 9,87E-10 | 1,51E-14 | 7,32E-06 |
| | **4.00** | 1,90E-08 | 1,58E-12 | 3,67E-05 | 3,40E-06 | 5,31E-09 | 6,29E-04 |
| | **5.00** | 5,41E-06 | 1,10E-08 | 8,14E-04 | 1,59E-04 | 2,12E-06 | 5,31E-03 |
| **103.00** | **1.00** | 2,33E-117 | 5,80E-185 | 1,97E-63 | 4,11E-65 | 3,76E-102 | 1,50E-35 |
| | **2.00** | 6,60E-31 | 5,56E-48 | 2,92E-17 | 9,48E-18 | 3,51E-27 | 3,39E-10 |
| | **3.00** | 8,83E-15 | 1,70E-22 | 1,35E-08 | 7,28E-09 | 3,50E-13 | 2,17E-05 |
| | **4.00** | 4,46E-09 | 1,62E-13 | 1,67E-05 | 1,07E-05 | 3,18E-08 | 1,18E-03 |
| | **5.00** | 2,11E-06 | 2,53E-09 | 4,84E-04 | 3,37E-04 | 6,77E-06 | 8,08E-03 |
| **104.00** | **1.00** | 1,39E-127 | 3,58E-201 | 6,95E-69 | 6,39E-58 | 9,34E-91 | 1,04E-31 |
| | **2.00** | 1,78E-33 | 4,77E-52 | 1,23E-18 | 6,22E-16 | 2,59E-24 | 3,23E-09 |
| | **3.00** | 6,22E-16 | 2,59E-24 | 3,23E-09 | 4,82E-08 | 6,79E-12 | 6,10E-05 |
| | **4.00** | 9,87E-10 | 1,51E-14 | 7,32E-06 | 3,17E-05 | 1,72E-07 | 2,16E-03 |
| | **5.00** | 7,93E-07 | 5,46E-10 | 2,83E-04 | 6,87E-04 | 2,04E-05 | 1,21E-02 |
| **105.00** | **1.00** | 3,06E-138 | 4,52E-218 | 1,44E-74 | 3,67E-51 | 4,76E-80 | 4,25E-28 |
| | **2.00** | 3,73E-36 | 2,76E-56 | 4,52E-20 | 3,19E-14 | 1,29E-21 | 2,71E-08 |
| | **3.00** | 3,93E-17 | 3,31E-26 | 7,30E-10 | 2,87E-07 | 1,11E-10 | 1,62E-04 |
| | **4.00** | 2,05E-10 | 1,28E-15 | 3,12E-06 | 8,84E-05 | 8,51E-07 | 3,83E-03 |
| | **5.00** | 2,87E-07 | 1,11E-10 | 1,62E-04 | 1,35E-03 | 5,77E-05 | 1,77E-02 |

**Table 4:** *Confidence bounds for the fraction nonconforming the lower and upper quality specification limits assuming a true value of the measurand of the control material equal to 100.00 measurement units, a sample of 20 control measurements, and a total allowable analytical error equal to 0.20.*

| confidence bounds for the fraction nonconforming at a 95% confidence level | | | | | | | |
|---|---|---|---|---|---|---|---|
| size of the sample of the measurements *n* = 40 | | | | | | | |
| total allowable analytical error *taae* = 0.20 | | | | | | | |
| **parameters of the sample** | | **fraction nonconforming the lower quality specification limit** | | | **fraction nonconforming the upper quality specification limit** | | |
| $\bar{X}$ | $s$ | **expected** | **lower bound** | **upper bound** | **expected** | **lower bound** | **upper bound** |
| **100.00** | **1.00** | 2,75E-89 | 3,94E-124 | 1,69E-59 | 2,75E-89 | 3,94E-124 | 1,69E-59 |
| | **2.00** | 7,62E-24 | 1,12E-32 | 2,79E-16 | 7,62E-24 | 1,12E-32 | 2,79E-16 |
| | **3.00** | 1,31E-11 | 1,27E-15 | 3,66E-08 | 1,31E-11 | 1,27E-15 | 3,66E-08 |
| | **4.00** | 2,87E-07 | 1,35E-09 | 2,89E-05 | 2,87E-07 | 1,35E-09 | 2,89E-05 |
| | **5.00** | 3,17E-05 | 9,02E-07 | 6,85E-04 | 3,17E-05 | 9,02E-07 | 6,85E-04 |
| **101.00** | **1.00** | 3,28E-98 | 1,31E-136 | 2,19E-65 | 8,53E-81 | 2,94E-112 | 6,76E-54 |
| | **2.00** | 4,32E-26 | 8,21E-36 | 9,09E-18 | 1,05E-21 | 1,08E-29 | 7,28E-15 |
| | **3.00** | 1,28E-12 | 4,98E-17 | 7,79E-09 | 1,20E-10 | 2,76E-14 | 1,60E-07 |
| | **4.00** | 7,60E-08 | 2,14E-10 | 1,19E-05 | 1,02E-06 | 7,79E-09 | 6,77E-05 |
| | **5.00** | 1,33E-05 | 2,74E-07 | 3,82E-04 | 7,23E-05 | 2,81E-06 | 1,20E-03 |
| **102.00** | **1.00** | 1,44E-107 | 1,07E-149 | 1,48E-71 | 9,74E-73 | 5,42E-101 | 1,40E-48 |
| | **2.00** | 1,91E-28 | 4,24E-39 | 2,52E-19 | 1,13E-19 | 7,37E-27 | 1,62E-13 |
| | **3.00** | 1,12E-13 | 1,68E-18 | 1,54E-09 | 9,87E-10 | 5,15E-13 | 6,52E-07 |
| | **4.00** | 1,90E-08 | 3,13E-11 | 4,70E-06 | 3,40E-06 | 4,13E-08 | 1,52E-04 |
| | **5.00** | 5,41E-06 | 7,87E-08 | 2,08E-04 | 1,59E-04 | 8,33E-06 | 2,04E-03 |
| **103.00** | **1.00** | 2,33E-117 | 2,18E-163 | 5,15E-78 | 4,11E-65 | 2,47E-90 | 1,51E-43 |
| | **2.00** | 6,60E-31 | 1,55E-42 | 5,92E-21 | 9,48E-18 | 3,55E-24 | 3,05E-12 |
| | **3.00** | 8,83E-15 | 4,88E-20 | 2,85E-10 | 7,28E-09 | 8,27E-12 | 2,48E-06 |
| | **4.00** | 4,46E-09 | 4,19E-12 | 1,79E-06 | 1,07E-05 | 2,01E-07 | 3,29E-04 |
| | **5.00** | 2,11E-06 | 2,14E-08 | 1,11E-04 | 3,37E-04 | 2,33E-05 | 3,40E-03 |
| **104.00** | **1.00** | 1,39E-127 | 1,09E-177 | 9,32E-85 | 6,39E-58 | 2,80E-80 | 8,44E-39 |
| | **2.00** | 1,78E-33 | 4,00E-46 | 1,18E-22 | 6,22E-16 | 1,21E-21 | 4,90E-11 |
| | **3.00** | 6,22E-16 | 1,21E-21 | 4,90E-11 | 4,82E-08 | 1,14E-10 | 8,77E-06 |
| | **4.00** | 9,87E-10 | 5,15E-13 | 6,52E-07 | 3,17E-05 | 9,02E-07 | 6,85E-04 |
| | **5.00** | 7,93E-07 | 5,52E-09 | 5,73E-05 | 6,87E-04 | 6,20E-05 | 5,53E-03 |
| **105.00** | **1.00** | 3,06E-138 | 1,36E-192 | 8,74E-92 | 3,67E-51 | 7,83E-71 | 2,45E-34 |
| | **2.00** | 3,73E-36 | 7,28E-50 | 2,01E-24 | 3,19E-14 | 2,92E-19 | 6,69E-10 |
| | **3.00** | 3,93E-17 | 2,58E-23 | 7,83E-12 | 2,87E-07 | 1,35E-09 | 2,89E-05 |
| | **4.00** | 2,05E-10 | 5,81E-14 | 2,29E-07 | 8,84E-05 | 3,71E-06 | 1,37E-03 |
| | **5.00** | 2,87E-07 | 1,35E-09 | 2,89E-05 | 1,35E-03 | 1,56E-04 | 8,77E-03 |

**Table 5:** *Confidence bounds for the fraction nonconforming the lower and upper quality specification limits assuming a true value of the measurand of the control material equal to 100.00 measurement units, a sample of 40 control measurements, and a total allowable analytical error equal to 0.20.*

**Figure 1:** *Confidence bounds for the fraction nonconforming the lower quality specification limit.*

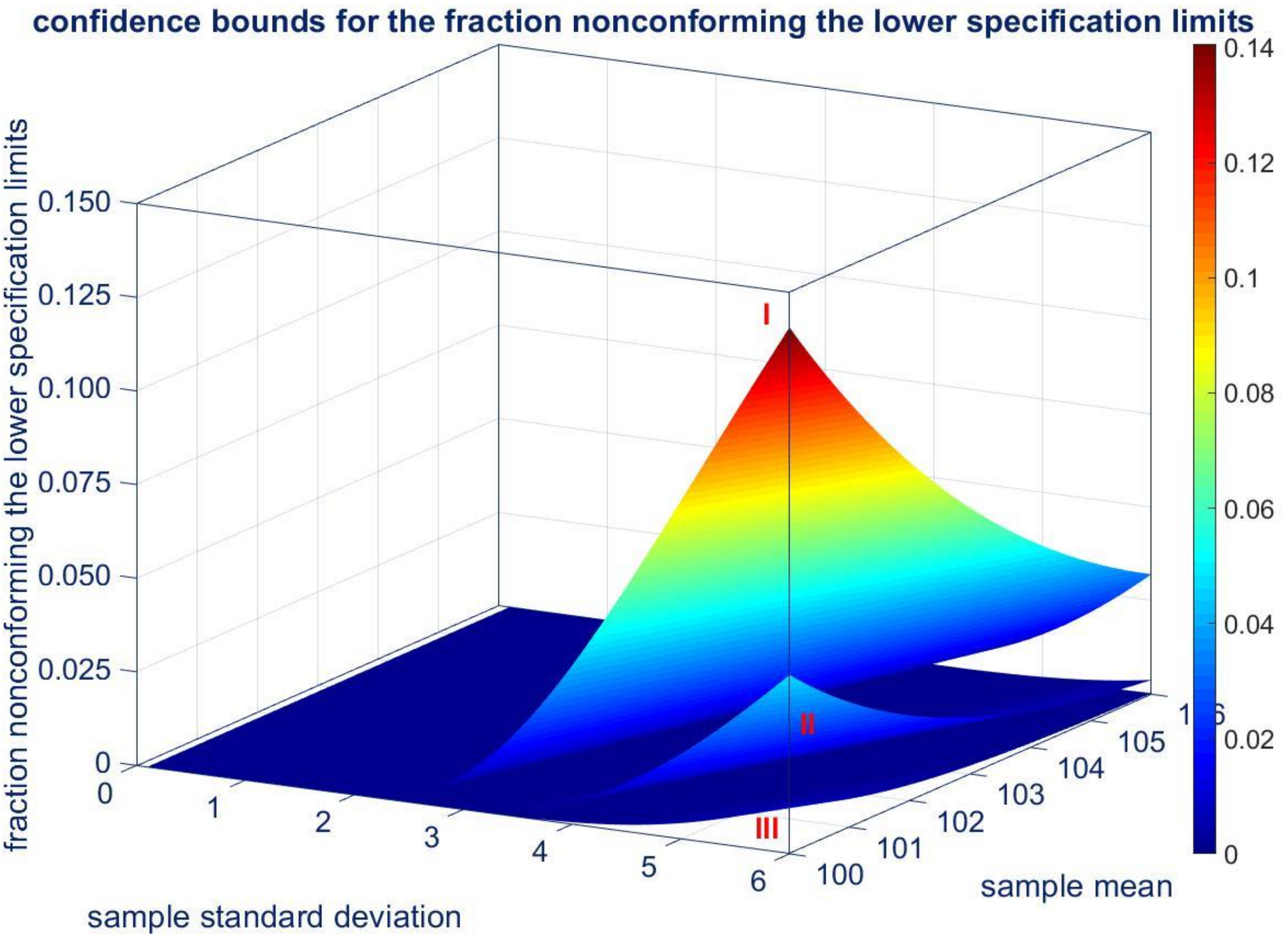

*The expected fraction nonconforming the lower quality specification limit (surface II) and the lower (surface III) and upper (surface I) confidence bounds for the fraction nonconforming at a 95% confidence level versus the sample mean and the sample standard deviation. It is assumed a known true value of the measurand of the control material equal to 100.00 measurement units, a sample of 20 control measurements, and a total allowable analytical error equal to 0.10.*

**Figure 2:** *Confidence bounds for the fraction nonconforming the upper quality specification limit.*

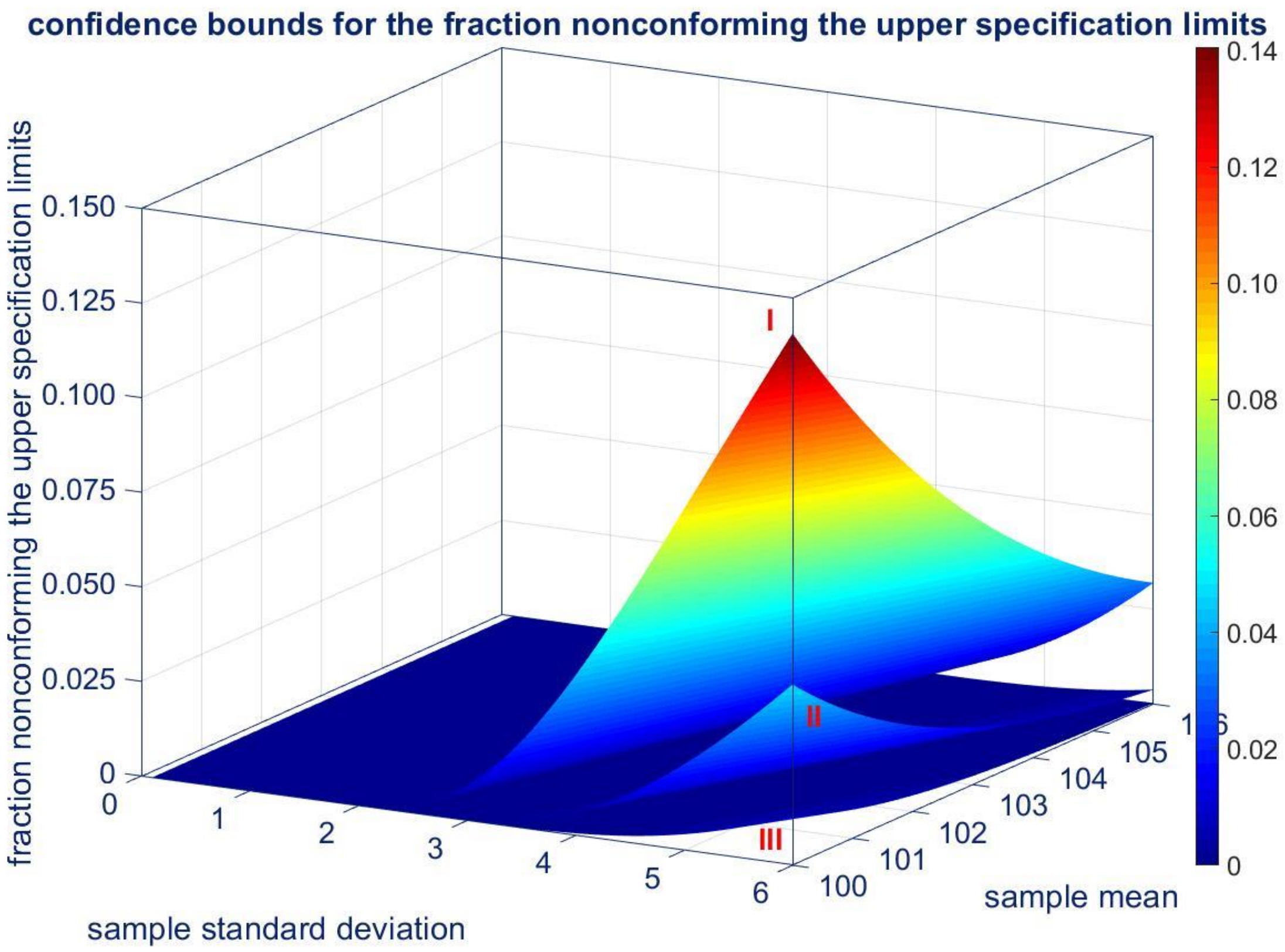

*The expected fraction nonconforming the upper quality specification limit (surface II) and the lower (surface III) and upper (surface I) confidence bounds for the fraction nonconforming at a 95% confidence level versus the sample mean and the sample standard deviation. It is assumed a known true value of the measurand of the control material equal to 100.00 measurement units, a sample of 20 control measurements, and a total allowable analytical error equal to 0.10.*

**Figure 3:** *Confidence bounds for the fraction nonconforming the lower quality specification limit.*

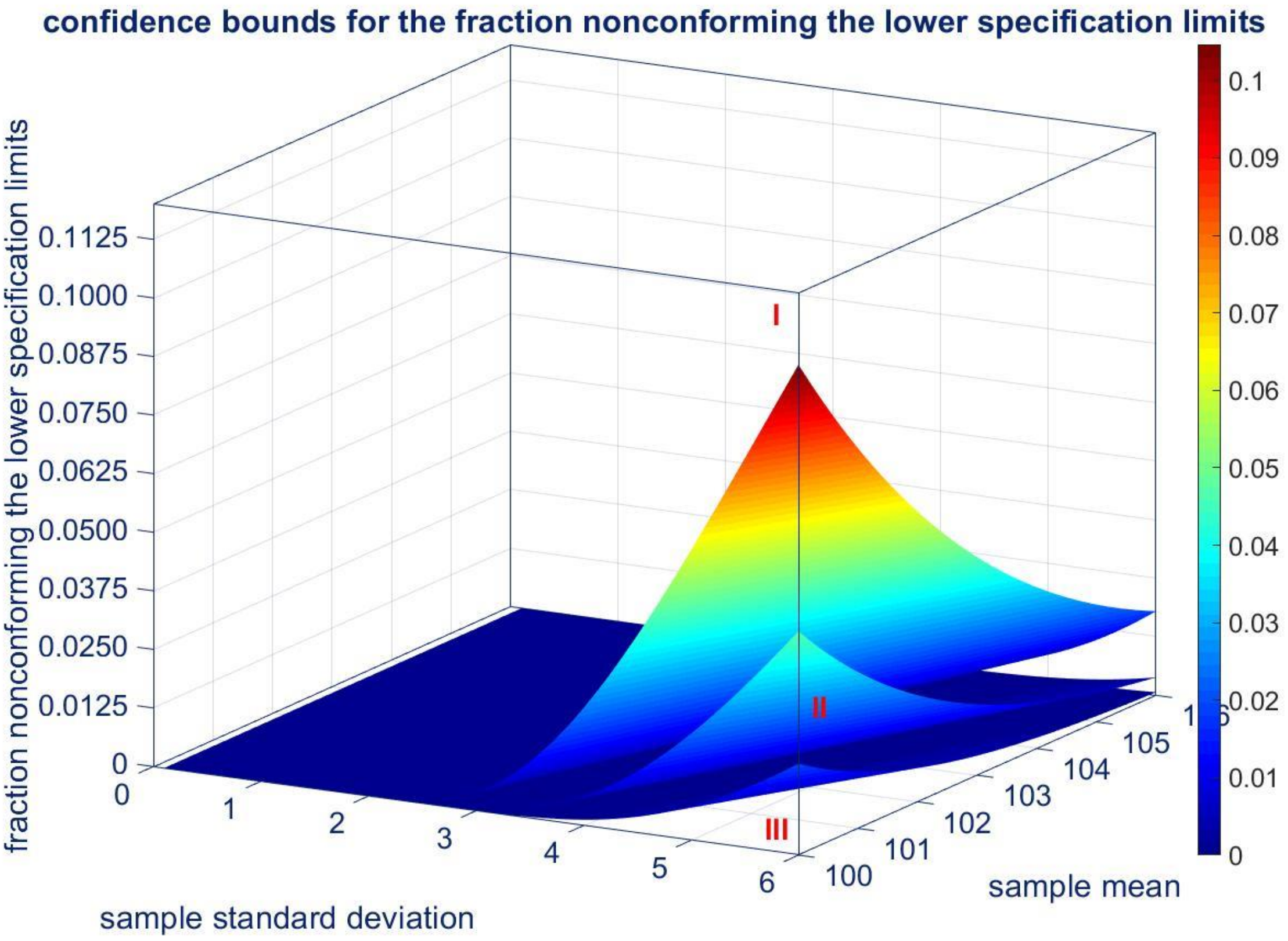

*The expected fraction nonconforming the lower quality specification limit (surface II) and the lower (surface III) and upper (surface I) confidence bounds for the fraction nonconforming at a 95% confidence level versus the sample mean and the sample standard deviation. It is assumed a known true value of the measurand of the control material equal to 100.00 measurement units, a sample of 40 control measurements, and a total allowable analytical error equal to 0.10.*

**Figure 4***: Confidence bounds for the fraction nonconforming the upper quality specification limit.*

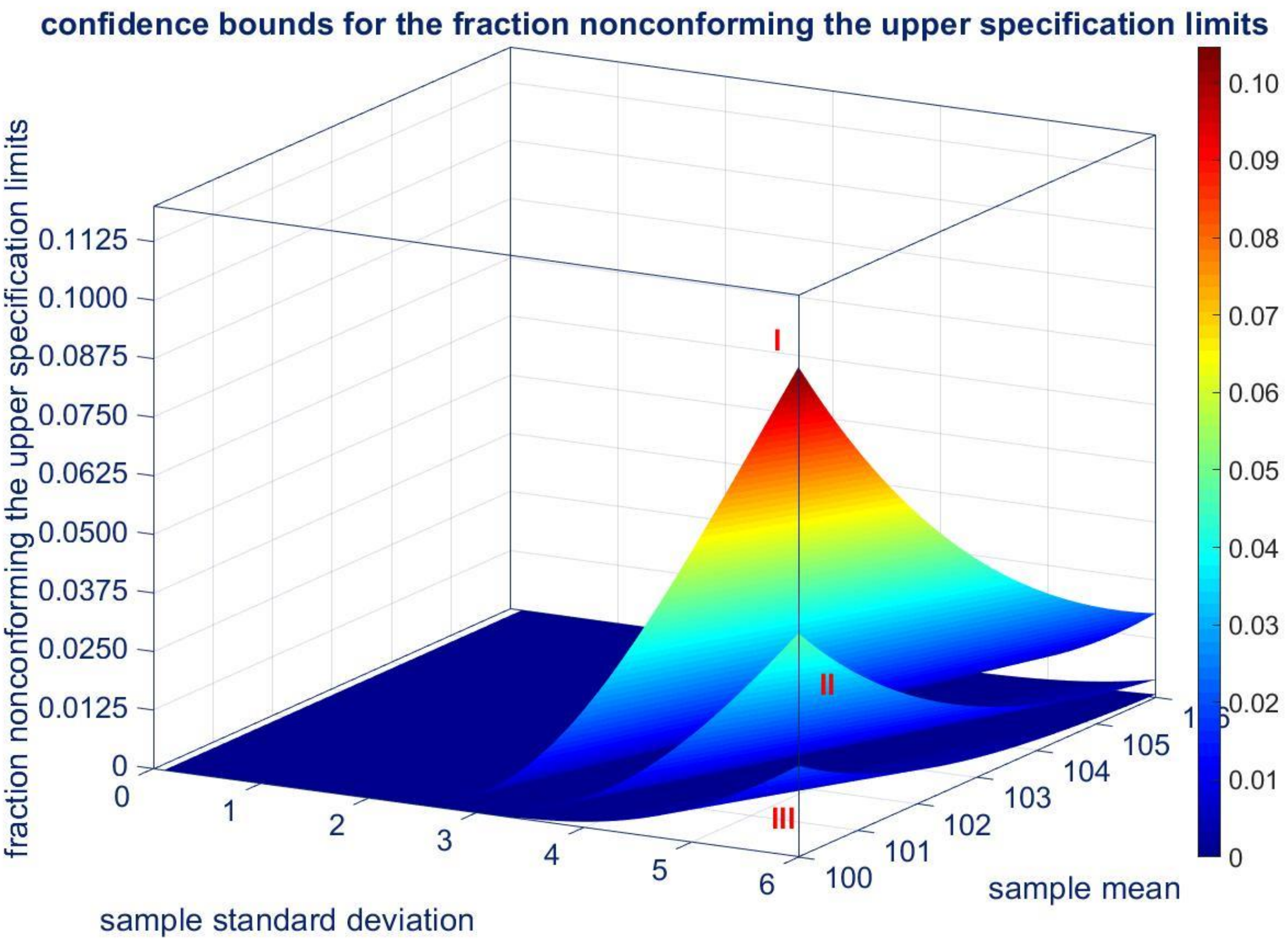

*The expected fraction nonconforming the upper quality specification limit (surface II) and the lower (surface III) and upper (surface I) confidence bounds for the fraction nonconforming at a 95% confidence level versus the sample mean and the sample standard deviation. It is assumed a known true value of the measurand of the control material equal to 100.00 measurement units, a sample of 40 control measurements, and a total allowable analytical error equal to 0.10.*

**Figure 5:** *Confidence bounds for the fraction nonconforming the lower quality specification limit.*

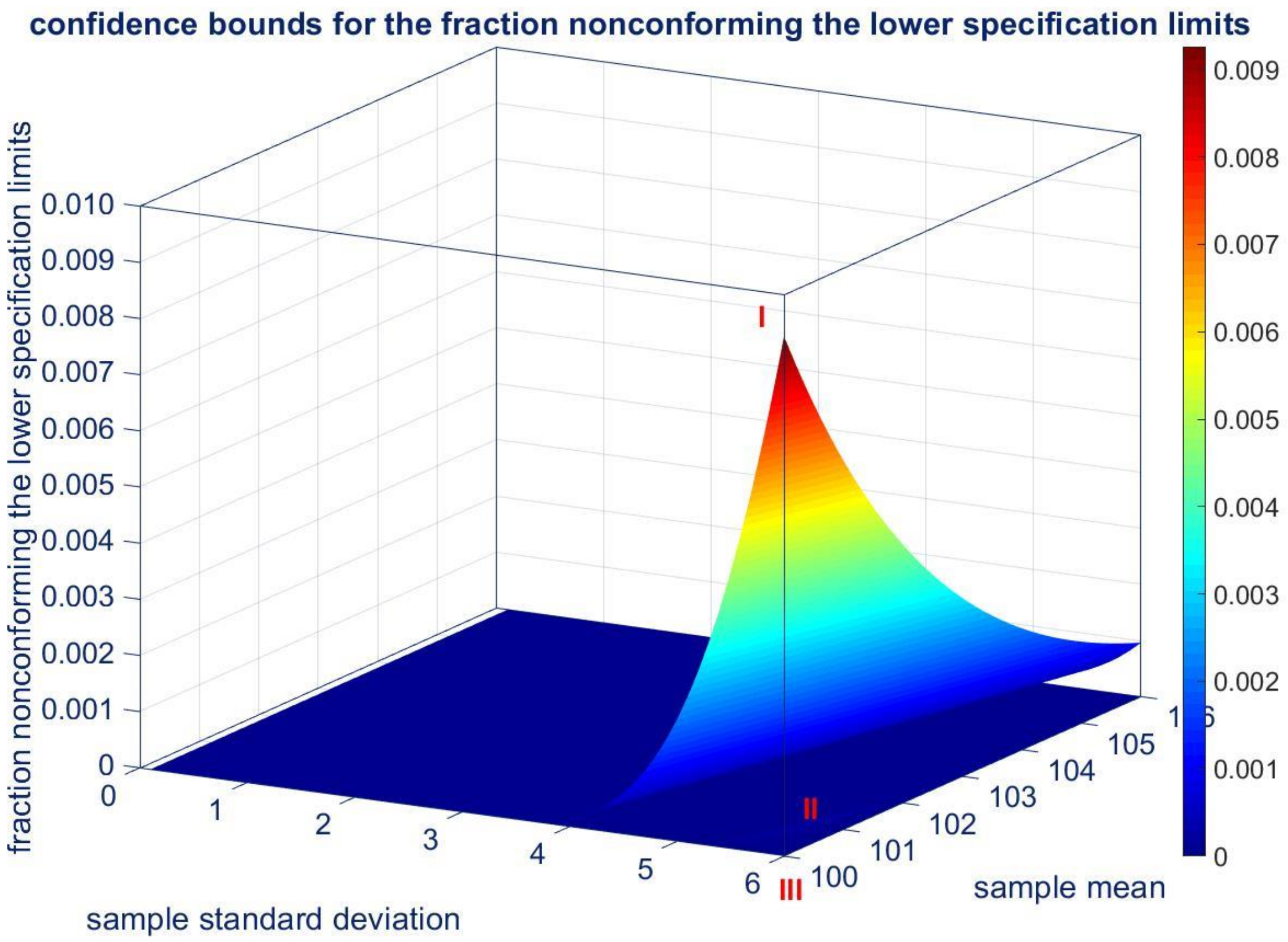

*The expected fraction nonconforming the lower quality specification limit (surface II) and the lower (surface III) and upper (surface I) confidence bounds for the fraction nonconforming at a 95% confidence level versus the sample mean and the sample standard deviation. It is assumed a known true value of the measurand of the control material equal to 100.00 measurement units, a sample of 20 control measurements, and a total allowable analytical error equal to 0.20.*

**Figure 6:** *Confidence bounds for the fraction nonconforming the upper quality specification limit.*

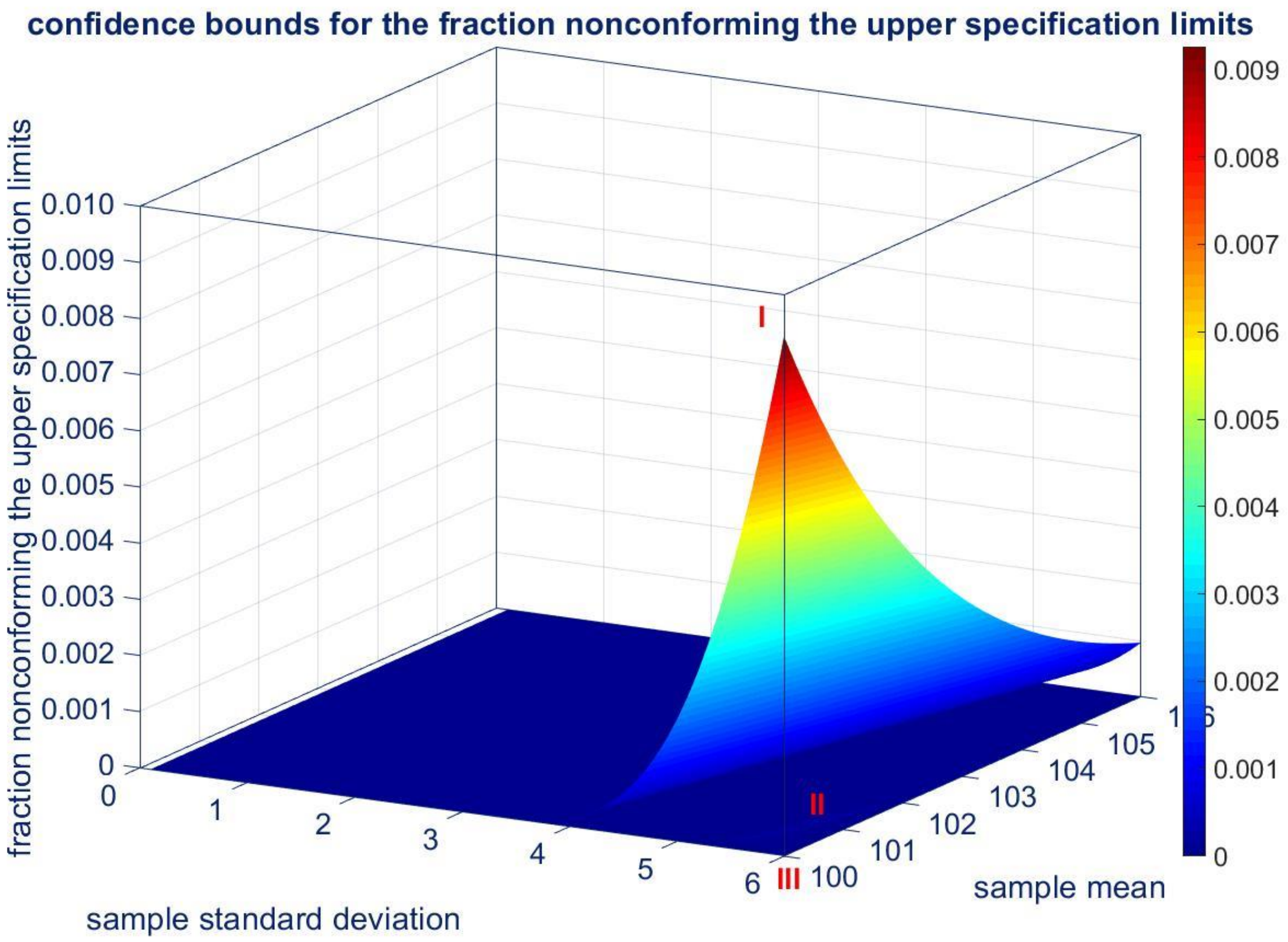

*The expected fraction nonconforming the upper quality specification limit (surface II) and the lower (surface III) and upper (surface I) confidence bounds for the fraction nonconforming at a 95% confidence level versus the sample mean and the sample standard deviation. It is assumed a known true value of the measurand of the control material equal to 100.00 measurement units, a sample of 20 control measurements, and a total allowable analytical error equal to 0.20.*

**Figure 7:** *Confidence bounds for the fraction nonconforming the lower quality specification limit.*

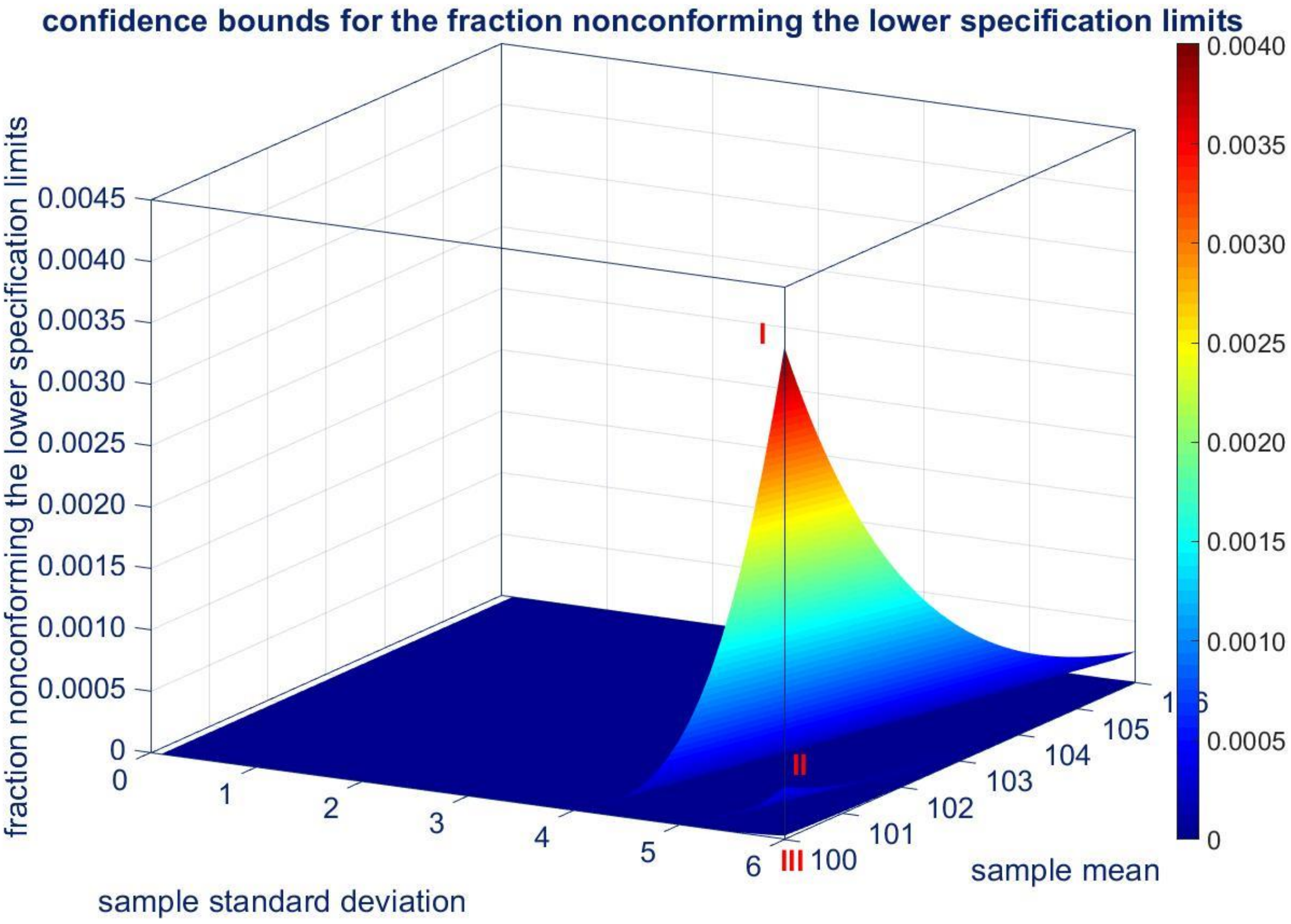

*The expected fraction nonconforming the lower quality specification limit (surface II) and the lower (surface III) and upper (surface I) confidence bounds for the fraction nonconforming at a 95% confidence level versus the sample mean and the sample standard deviation. It is assumed a known true value of the measurand of the control material equal to 100.00 measurement units, a sample of 40 control measurements, and a total allowable analytical error equal to 0.20.*

**Figure 8:** *Confidence bounds for the fraction nonconforming the upper quality specification limit.*

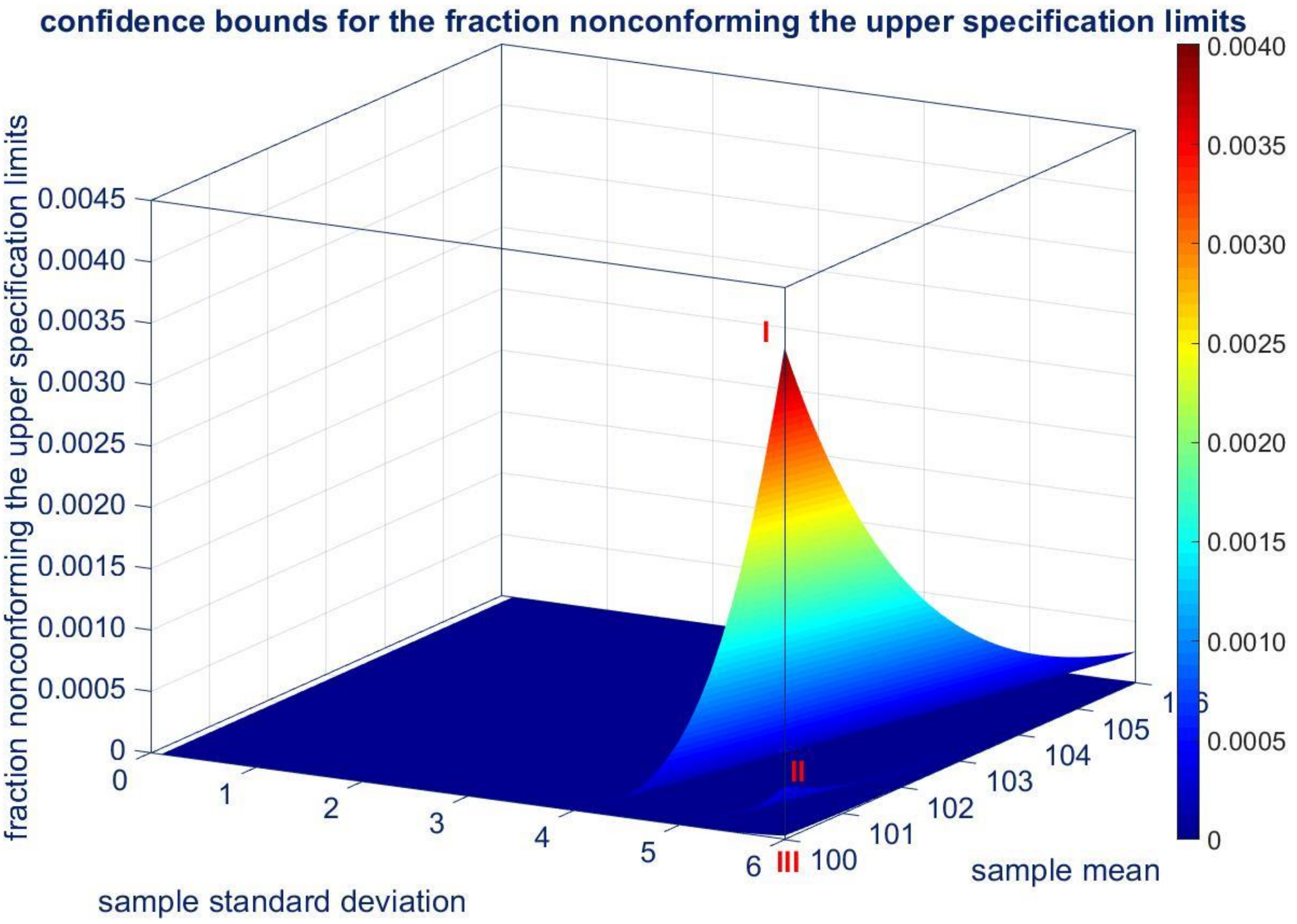

*The expected fraction nonconforming the upper quality specification limit (surface II) and the lower (surface III) and upper (surface I) confidence bounds for the fraction nonconforming at a 95% confidence level versus the sample mean and the sample standard deviation. It is assumed a known true value of the measurand of the control material equal to 100.00 measurement units, a sample of 40 control measurements, and a total allowable analytical error equal to 0.20.*

## 5. Discussion

In clinical laboratory medicine the fraction nonconforming besides being a quality measure related to various process capability indices, it is used for the definition of the critical errors during the QC design process. To estimate the fraction nonconforming the upper and lower quality specification bounds for the measurement error are needed. These are defined by the maximum total clinically allowable analytical error (Linnet 1989).

As both the systematic and the random error of any measurement procedure are actually unknown, we can calculate the respective confidence bounds for the fraction nonconforming each of the upper and lower quality specification limits, using the noncentral *t*-distribution. The calculation is mathematically complicated and computationally intensive. It is not possible to calculate the confidence bounds for the total fraction nonconforming either the upper or the lower quality specification limits, without additional assumptions (Owen 1965).

The results of the study show that the difference of the confidence bounds for the fraction nonconforming from the expected fraction nonconforming is considerable for greater random and systematic errors and for smaller sample sizes, even when we assume an insignificant uncertainty of the assigned value of the measurand of the control material. Consequently it is considerable when the expected fraction nonconforming equals its acceptable upper bound that is usually set equal to 0.10. Therefore, the uncertainty of the calculation of the critical errors is considerable as well, as Gernand and Wojciechit have pointed out previously (Gernand 2006). Accordingly, the uncertainty of the whole quality control design process is significant, even assuming a negligible uncertainty of the assigned value of the measurand of the control material.

Although it was assumed a reference control material with a measurand with an assigned value of 100.00, the tables and figures of this study can be used for the estimation of the confidence bounds for the fraction nonconforming for any measurand with any assigned value.

## 6. Conclusion

The confidence bounds for the fraction nonconforming are useful quality measures, as the expected fraction nonconforming is a key measure of the quality design process. The confidence bounds for the fraction nonconforming each of the upper and lower quality specification limits can be calculated using the noncentral *t*-distribution.

## 7. Acknowledgements

I am deeply grateful to Dr. Fritz Scholz for providing me with the technical report on the applications of the noncentral t- distribution (Scholz 1994), that was the basis of this technical report, as well as for his suggestions and corrections.